\documentclass[usenatbib,onecolumn,useAMS]{mn2e}
\usepackage{epsfig}

\def\msun{{\rm M_{\odot}}}

\title [Disc fragmentation ]
{Pseudo-viscous modeling of self-gravitating discs and the formation
of low mass ratio binaries}
\author[C.J.~Clarke]{C.J.~Clarke$^1$\\
$^1$Institute of Astronomy, Madingley Rd, Cambridge, CB3 0HA, UK}
\date{Submitted: July 2103}

\pagerange{\pageref{firstpage}--\pageref{lastpage}} \pubyear{2003}

\begin{document}
\def\lta{\mathrel{\spose{\lower 3pt\hbox{$\mathchar"218$}}
     \raise 2.0pt\hbox{$\mathchar"13C$}}}
\def\gta{\mathrel{\spose{\lower 3pt\hbox{$\mathchar"218$}}
     \raise 2.0pt\hbox{$\mathchar"13E$}}}
\def\Msun{{\rm M}_\odot}
\def\msun{{\rm M}_\odot}
\def\Rsun{{\rm R}_\odot}
\def\Lsun{{\rm L}_\odot}
\def\19{GRS~1915+105}
\label{firstpage}
\maketitle

\begin{abstract}
We present analytic models for the local structure of self-regulated self-gravit
ating accretion discs that are subject to realistic cooling. Such an approach
can be used to predict the secular evolution of self-gravitating discs (which
can usefully be compared with future radiation hydrodynamical simulations)
and
to define various physical regimes as a function of radius and equivalent steady
 state accretion rate.
We show that fragmentation is  inevitable, given realistic rates
of infall into the disc, once the disc extends to radii $> 70$ A.U. (in the
case of a solar mass central object). Owing to the outward redistribution
of disc material by gravitational
torques, we also predict fragmentation at $> 70$ A.U. even in the
case of low angular momentum cores which initially collapse to a much
smaller radius. We point out that $70$ A.U. is close to the median binary
separation and propose that such delayed fragmentation, at the point
that the disc expands to $> 70$ A.U., ensures the creation of low mass ratio
companions that can avoid substantial further growth and
consequent evolution towards unit mass ratio.
We thus propose this as a promising mechanism for
producing low mass ratio binaries, which, while abundant observationally,
are severely underproduced
in hydrodynamical models. 
\end{abstract}


\section{Introduction}

 Since the early attempts of Paczynski 1978, Lin \& Pringle 1987, there has been
considerable interest in describing the action of self-gravity
in accretion discs through adaption of the standard formulation
of viscous disc theory. In such an approach, the net effect of
self-gravitating modes in discs is to provide a phenomenological
viscosity, in much the same way that the action of the magneto-rotational
instability (MRI) is often discussed as a source of a  pseudo-viscosity.
Indeed, self-gravity is a prime candidate `viscosity mechanism' in
the early stages of evolution of protostellar discs, both because
the MRI is unlikely to be  efficacious in dense regions of
cool, weakly ionised discs  (Gammie 1996) and
also because it is hard to avoid the conclusion - in a scenario where
the star forms from material channeled by the disc - that at early times
the disc's {\it self} -gravity should be important.

  There are obvious advantages of being able to model the effect of disc
self-gravity (or indeed, the MRI) using a pseudo-viscous prescription,
most notably the
fact that one can use many elements of the apparatus of
`$\alpha$'  accretion disc theory (Shakura \& Sunyaev 1973)
to compute the structure and
evolution of self-gravitating discs. For example, one may compute
the structure of a steady state disc (as a function of accretion rate:
see Rakikov 2009)
or else, given a disc's  surface density and temperature distribution, can
evaluate the disc's evolution  using  the viscous diffusion
equation.

  The exploration of this approach has been encouraged by
some
obvious superficial similarities
between the action of viscous torques and those produced by self-gravitating
features in the disc. Non-axisymmetric modes produce torques that
redistribute angular momentum within the disc and the work done by these
torques is dissipated in the disc, providing a net conversion of mechanical
energy into heat which permits the driving of an accretion flow.
In these respects, therefore, the action of self-gravity is
analogous to that of viscosity and it is tempting to describe it
as such. In this case, the only difference between a self-gravitating
disc and a conventional viscous disc is that the magnitude of the
gravitational pseudo-viscosity is not known {\it a priori}
(cf Bertin 1997), but instead self-adjusts
in order to keep the disc in a state of marginal gravitational stability.
Such a picture is again consistent with a range of simulations of 
self-gravitating
gas discs, which demonstrate such self-regulation  (e.g. Lodato \& Rice 2004,
Boley et al 2006, Mayer et al 2007, Stamatellos et al 2007, Cossins et al 2009):  the
amplitude of spiral features
self-adjusts so as to heat the disc to a point of marginal gravitational
stability (specifically such that the Toomre Q parameter - see Section
2.1 below -
is of order unity).

  However, this broad similarity between some of the effects of
viscosity, and of self-gravitating modes, is
not in itself sufficient for one to adopt a pseudo-viscous description:
in addition, one also needs a more precise requirement to
be satisfied, namely that, as in the viscous situation,
the rate of work done on the
flow locally is entirely specified by the local torque.
As pointed out by
Balbus \& Papaloizou 1999, there is an alternative situation where the
energy extracted from the flow is not necessarily dissipated {\it locally}
but is instead transported in a propagating wave, to
be dissipated at some other radial location. Whereas this does not
alter the global energy balance (i.e. the rate of mechanical energy
lost by the accretion flow is still equal to the total rate of
energy dissipation, integrated over the disc) it obviously prevents
a simple relationship between local torques and the local thermodynamic
state of the disc (see Lodato \& Bertin 2001 for a discussion of how
such transport could affect the spectrum generated by self-gravitating
discs).

  Such global energy transport is important in the case that
waves, which are launched at corotation resonances, are able to propagate
over significant radial distances. Thus a measure of the importance
of non-local effects is provided by how far waves persist away from
co-rotation. Balbus \& Papaloizou were thus able to compute the
importance of global effects via an `anomalous flux' which depends on
$(\Omega - \Omega_p)/\Omega$, the fractional deviation between the local
flow speed and the mode's pattern speed. However, one has to rely on numerical
simulations in order to discover the spectrum of modes (and their
pattern speeds) that are excited in the disc and thus cannot evaluate
the importance of this effect {\it a priori}.

  Numerical simulations however also allow one to test the pseudo-viscous
hypothesis directly, by measuring both the torques in the disc and the
local energy dissipation rate, and then comparing this relationship with
that expected in the case of local dissipation.
(Historically, this was first undertaken by Gammie 2001, although, as
stressed by Balbus \& Papaloizou, the form of the boundary conditions
in shearing box simulations ensures locality in any case).
More notably, a variety of global simulations, both SPH and grid
based calculations, show  that the relationship between torques and
energy dissipation rate is indeed similar to what would apply in the
viscous case (Lodato \& Rice 2004,2005, Boley et al 2006). This  essentially local behaviour
appears to be set by the requirement that the flow speed of the gas into
spiral arms is marginally supersonic, which prevents waves from propagating
far from corotation (Cossins et al 2009).  Such an argument implies that in more massive discs,
which are hotter in a state of marginal gravitational stability, the
higher sound speed should allow waves to propagate further from corotation,
and that  non-local effects are expected to be important. Simulations
corroborate this tendency, in that deviations from local behaviour
are more pronounced for more massive discs (Lodato \& Rice 2005,
Cossins et al 2009). Local energy dissipation however appears to be
an adequate approximation in the case of discs with masses up to several
tenths 
of the central object mass (Lodato \& Rice 2005, Cossins et al 2009).

  The above discussion suggests that in the case of self-gravitating
protostellar discs,
with $M_d/M_* < 0.5$, it is adequate to model the effects of self-gravity
in pseudo-viscous terms, and thus reap all the benefits of being able
to derive both steady state disc structures as well as the secular
evolution of the  disc. This latter is not currently feasible
in the case of hydrodynamic simulations: such simulations are typically
run for hundreds of disc outer orbital timescales, but the torques
measured in these simulations imply secular evolution timescales that
exceed this by at least an order of magnitude. Likewise, it is only
possible to model self-gravitating discs over a rather limited
range of parameter space: if the spiral modes are too strong, the
disc fragments (Gammie 2001, Rice et al 2005), whereas if they are too weak, the angular momentum
transfer in the disc is dominated by numerical viscosity. In practice,
this means that it is possible to perform hydrodynamic modeling over
a range of mode strengths which, when parameterised in terms of an
equivalent viscous $\alpha$ value (Shakura \& Sunyaev 1973), span only
about a factor three in $\alpha$. As we will see below, the $\alpha $
values that are expected in discs with realistic cooling regimes and
mass input rates instead span many orders of magnitude.

 The structure of the paper is as follows. In Section 2, we set out an
analytic description of 
self-regulated, self-gravitating discs in various cooling regimes, discussing
also the range of parameter space for which this description applies.
In Section 3 we discuss the resultant solutions for disc properties as
a 
function of steady state accretion rate
and radius. In Section 4, we set out how such solutions can be used also
to calculate the secular evolution of discs formed from collapsing cores,
which can be usefully compared with hydrodynamic solutions. We also apply
our results to binary star formation and point out that the use of
realistic disc cooling offers the prospect of being able to create binaries
with low mass ratio, an outcome that has 
an outcome that has been under-represented
in previous hydrodynamical models that do not incorporate such effects.
Section
5 recapitulates our main conclusions.

\section{Self-regulated self-gravitating discs }
\subsection{General description}

 The importance of self-gravity in discs is measured by the Toomre Q parameter,
which, in
the case of discs which are considerably less massive than their central
object and where the  rotation curve is thus approximately Keplerian, takes
the form:

\begin{equation}
Q = {{c_s \Omega}\over{\pi G \Sigma}}
\end{equation}

Here $\Sigma$ is the disc surface
density, $\Omega$ is the angular frequency and $c_s$ is the sound speed. 
  Formally, the condition $Q=1$ marks the stability boundary of discs
to {\it axisymmetric} instability (Toomre 1964). More generally, however, it is found
numerically that self-gravitating discs that are subject to cooling
evolve to a marginally unstable
self-regulated state with $Q$ in the range $1-2$. 
This result may be readily explained in heuristic
terms: gravitational collapse is opposed on small scales by pressure
and on large scales by shear. If $Q > 1$, there is an overlap between
the regimes of spatial scale that are stabilised by each of these
effects, and thus collapse cannot occur on any scale. Self-gravity
plays a minor role in discs with $Q >>1$, whereas discs with
$Q \sim 1$, which will play an important role in our subsequent
discussions, are in a state of marginal stability.
  
  Since the sound speed appears in the numerator of the definition
of $Q$, the fate of a disc set up with a given surface density
profile is largely set by its temperature, and hence on the
thermal equilibrium that it attains between radiative cooling
and the dissipation of mechanical energy (or any other source
of heating such as external irradiation). Internal dissipation is  
associated with the redistribution
of mass and angular momentum in the disc resulting from internal torques, 
which may be either viscous (e.g. stresses associated with the
magnetorotational instability) or gravitational
in origin. In the latter case, therefore, the development of self-gravitating
modes can provide an effective heat source which acts so as to stabilise 
the disc. If this heating however increases $Q$ significantly above
unity, then the modes themselves shut off, the disc cools and   the
modes begin to develop again. In this way, the disc achieves
a state of {\it self-regulation} where $Q$ is maintained close to unity.
\footnote{A similar mechanism sustains stellar discs that
are subject to `cooling' in the marginally
stable state, except here gravitational instability feeds energy
into non-circular motions of the stars rather than the internal energy
of the fluid and `cooling' consists of re-supply of stars on
circular orbits; Sellwood and Carlberg 1984.}
 
 The above discussion therefore implies that the setting up of a
disc in the gravitationally unstable regime does not {\it necessarily}
lead to fragmentation, since an alternative route is the establishment
of a self-regulated state of marginal gravitational stability. 
The important work of Gammie established the criteria that determine
which route a disc takes in practice: through local (shearing box)
simulations of self-gravitating discs subject to cooling on a
a timescale $t_{cool}$, he demonstrated that the boundary between
fragmentation and self-regulation is set by the requirement
$t_{cool} = 3 \Omega^{-1}$. Similar delineation of the fragmentation
boundary in terms of the ratio of cooling to dynamical timescale has been
found in a range of other (global) simulations (e.g. Rice et al 2005,
Clarke et al 2007), though the exact location of this boundary
depends on the dimensionality of the simulations. A fragmentation
boundary of this type may be qualitatively
understood inasmuch as fragmentation requires that the pdV work
done on overdense regions is radiated away on the (roughly dynamical)
timescale on which perturbations grow.

  A critical value of the cooling timescale can also be understood,
in the framework of the pseudo-viscous description of self-gravitating
discs, in terms of a critical value of the stress in the disc. As
noted in Section 1, a range of simulations have shown that the
relationship between the ${R,\Phi}$ component of the
stress tensor and the local dissipation rate is roughly the same
as that expected for a viscous process; in thermal equilibrium, this
then implies (in the case of a Keplerian disc) the relationship:

\begin{equation}
t_{cool} \sim {{4 \Omega^{-1}}\over{9 \gamma (\gamma-1) \alpha}}
\end{equation}

where $\gamma$ is the  ratio of specific heats and $\alpha$ is
the usual  parameterisation of the ${R,\Phi}$ component of the 
stress tensor, ($W(R,\Phi)$ ), as a fraction of the thermal
pressure
(Shakura and Sunyaev
1973). Thus a given ratio of $t_{cool}/\Omega^{-1}$ also reflects
a given value of $\alpha$ (at fixed $\gamma$). Rice et al
furthermore delineated the fragmentation 
boundary as a function of the ratio of specific heats, $\gamma$, 
and established that 
the fragmentation boundary actually
corresponds to a fixed maximum $\alpha$ value 
($\sim 0.06$).

\subsection{Calculation of disc structure}

 We set out below a self-gravitating disc model which
assumes a) self-regulation of the Toomre Q parameter to a value
$\sim$  unity and b) local thermal and hydrostatic equilibrium of the
disc under the assumption that the dissipation of energy associated with
gravitational modes can be modeled as a pseudoviscous process. 
We therefore parameterise the effect of self gravity as an agent for
angular momentum transport in terms of the conventional viscous
$\alpha$ parameter, though stress that, since this is no more than
a convenient parameterisation, we make no assumption that this
$\alpha$ is spatially constant (cf Bertin \& Lodato 1999). We also recognise
that since self-regulated discs are
typified by transient but regenerative spiral features, 
the local $\alpha$ values derived from simulations 
undergo
considerable fluctuations, and thus that the $\alpha$ values
that we derive here should correspond to time averages over
several orbital periods. {\footnote {The secular evolution timescales
that correspond to the $\alpha$ values that we derive are many
orders of magnitude greater than the orbital timescale, so that such
averaging is appropriate}}.

  We will find that there are several regions of parameter space to
which the model described above does not apply, because some other
agent is providing energy input (and possibly angular momentum transport).
Thus, for example, the disc may be non-self gravitating (with angular
momentum transport provided by the MRI) or else (although self-gravitating
and subject to associated angular momentum transport) may have its
thermal properties mainly set by external irradiation. 
 We therefore  clarify the self-consistency checks
that we run before assigning a solution to the  
`standard self-gravitating' category.

\subsection{Self-consistency checks} 

  We will find that over much of the relevant parameter space
(in terms of radius and steady state accretion rate), the disc
exists in a  self-gravitating state with thermal equilibrium
between heating associated with dissipation of energy contained in 
gravitational modes and radiative cooling.  We however also
allow other possibilities by running three checks on the resulting
solutions: (i) we check that the effective value of $\alpha$ for
the solution does not exceed $\alpha_{max} \sim 0.06$ and, if it does,
designate this a fragmenting solution (see Section 2.1 above), 
(ii) we check that the
equilibrium temperature does not fall below a minimum value
$T_{min}$ which, following Hartmann et al 1998, we set to
$10$K as representing the minimum temperature that gas can
attain when subject to ambient heat sources in molecular clouds.
We designate such solutions as isothermal and replace the thermal
equilibrium condition by imposing $T=T_{min}$, (iii) we check the
effective value of $\alpha$ for
the solution in order to see whether it is less than the value of
$\alpha$ which we would expect from the operation of the MRI in this
region of the disc. This check therefore involves both an assumption
about the region of the disc that is potentially subject to the
MRI and about the typical
value of $\alpha$  ($\alpha_{MRI}$) delivered by the MRI. Both these quantities
are rather uncertain. In the former regard, we need the ionisation
fraction to exceed  $10^{-13}$ (Gammie 1996), 
either as a result of thermal ionisation, in
warm inner disc regions, or through ionisation (by cosmic rays or
Xrays, Glassgold et al 1997) in the outer (low column density) regions of the disc
. 
We thus assume that the
MRI may be  activated either if the mid-plane temperature exceeds
$1000$ K or at a radius $ > R_{dead}$. In reality, 
$R_{dead}$ should itself be a function of accretion rate; since
a full investigation of this
effect is beyond the scope of this paper, we instead choose a fixed
value $R_{dead} \sim 100 A.U.$. This is towards the upper range of
values quoted in the literature (Sano et al 2000, Fromang et al 2003,
Matsumura \& Pudritz 2003), which we justify by the fact
that we are mainly focusing on rather massive discs where $R_{dead}$ is
likely to be higher. 
The value of $\alpha$ delivered
by the MRI is a fluctuating quantity, whose time average has been
reported over a wide range, mainly in the domain $< 0.01$ (see discussion
in King et al 2007); here we adopt $\alpha_{MRI}=0.01$.
We note that our analytical solutions can be readily
generalised to other choices of $r_{dead}$,  $\alpha_{MRI}$
and $T_{min}$ (see Rafikov 2009).

  If $T < 1000$K and if $R<R_{dead}$ then we assume that there is
no possibility of the MRI being activated (in other words, we do not
here consider the possibility of layered accretion or else of
activation of the dead disc mid-plane by interaction with the
surface active layer; Fleming \& Stone 2003). Since we are only
considering two sources of angular momentum redistribution
(i.e. the MRI and self-gravity) then this implies that the
disc {\it must} be self-gravitating in this regime if it is also
accreting. In other regions of the disc, which are 
MRI active in principle, we then have to check  whether the resulting MRI
solution
(at given accretion rate and radial location) is self-consistent
(in the sense of being non-self gravitating).  
If instead the Toomre $Q$ parameter corresponding to this MRI solution
is $ < 1$, we infer that the angular momentum transport is dominated
by self-gravity and revert to the standard self-gravitating solution. 
We note that the interface between the MRI and self-gravitating
regimes   
corresponds to $\alpha = 0.01$, $Q=1$ and that therefore once the disc
becomes self-gravitating (in regions where the MRI should be active in
principle) then the corresponding $\alpha > 0.01$. Since we deem
that the disc fragments once $\alpha > \alpha_{frag} = 0.06$, this implies that even in
parts of the disc that are potentially MRI active, there
is a wedge of parameter space  (for which  $\alpha$ is in the range $0.01-0.06$)
where the disc is self-gravitating and
self-regulated.  We note that the existence of such  a region is a consequence
of the fact that $\alpha_{MRI}$ is somewhat below $\alpha_{frag}$. 
Finally, we note that there is a  wedge of parameter space for which
we cannot find self-consistent steady state solutions using either the
MRI or self-gravity as angular momentum transfer agents - in this case the
self-gravitating solution has $T> 1000$K and so the MRI should be active
in this region, delivering a value of $\alpha$ that is higher than the
corresponding self-gravitating value. However, if one instead seeks an
MRI solution at this location, 
the higher $\alpha$ value implies a cooler disc (at given accretion
rate) so that the temperature is now $<1000$K and the MRI should not be
activated. We denote regions of parameter space for which this is the
case by the ??? in Figures 1-6 and note that in practice this will
cause time-dependent behaviour, as envisaged by  Gammie 1999,
Armitage et al 2001.

\subsection{The standard self-gravitating regime}

 Given a location in the disc, radius $R$, and local surface
density, $\Sigma$, the self-regulated condition $Q=1$ (equation
(1)) immediately fixes the local temperature, $T$. Given $T$
and $\Sigma$, hydrostatic equilibrium yields the disc scale
height through the standard thin disc relation $H = c_s/\Omega$
and the mid-plane density $\rho = \Sigma/2 H$. 
\footnote {This
expression for $H$ is appropriate to the case that the
main vertical component of the gravitational
force is provided by the central star (or disc at much smaller
radius), rather than the disc's gravity locally. Actually, 
the two effects are competitive with each other for a disc with
$Q \sim 1$, 
and we have therefore
over-estimated $H$ (and underestimated $\rho$) only slightly.
See Bertin \& Lodato 1999 for expressions approximating $H$ in the
self-gravitating regime.}  

We then use 
the power law fits to  the Rosseland mean opacity
$\kappa(\rho,T)$ contained in Bell and Lin (1994)
in order to compute the optical depth ($\tau= \kappa \Sigma$)
and hence to determine whether the disc is locally
optically thick or optically thin. As it turns out that the disc
is always optically thick in the regime $T> T_{min}$,
we  calculate the
(one-sided) local cooling rate per unit area as:

\begin{equation}
Q^- = {{8 a c T^4}\over{3 \tau}}
\end{equation}



(Note that is effectively a one zone model vertically, inasmuch as
it calculates the optical depth using opacity values characteristic
of the mid-plane temperature; since the disc photosphere is cooler than
the mid-plane, this will not be accurate in regions where the
opacity is  highly temperature
dependent). 

  In order to calculate the effective viscosity locally, we then
apply the thermal equilibrium condition:

\begin{equation}
Q^+ = Q^-
\end{equation}

where $Q^+$ is the local rate of energy dissipation 
due to the effective viscosity, $\nu$:

\begin{equation}
Q^+ = 9/8 \nu \Sigma \Omega^2
\end{equation}

 Hence we have calculated the effective kinematic viscosity
($\nu(\Sigma,R)$) in the standard self-gravitating regime, and
could use this to evolve the disc by solving the time-dependent 
viscous diffusion
equation. We can also determine the
effective $\alpha$ value corresponding to this $\nu$  via
the definition:

\begin{equation}
\alpha = {{3 \nu  \Omega}\over{ 2 c_s^2}}
\end{equation}

Comparison of  this value with $\alpha_{frag}$ and
$\alpha_{MRI}$ then establishes whether the solution is instead
in the MRI or fragmenting regime. In the former case (i.e. where
$\alpha < \alpha_{MRI}$), then provided
that the disc is potentially MRI active for this value of $\Sigma$ and
$R$ (see 2.3 above), 
the condition $Q=1$ is relaxed and is replaced by the restriction
$\alpha = \alpha_{MRI}$ (i.e. we compute standard fixed
$\alpha$-disc solutions, cf Bell and Lin 1994). If, conversely,
$\alpha > \alpha_{frag}$ then the disc is deemed to be subject to
fragmentation and is labelled as such in Figures 1-6.     

\subsection{Caveats}
The association of fragmentation with a given value of $\alpha$ (or
cooling timescale) is mainly based on simulations that set up discs
with simply parameterised cooling at a variety of rates. There
are two possible objections to this approach. Firstly, it is not
immediately clear that the same fragmentation boundary would apply in the
more realistic case where a state of rapid cooling is approached on a slow
(secular) evolution timescale. Clarke et al 2007 however did not find that
the location of the fragmentation boundary was unduly sensitive to the
rate at which these conditions were approached, implying that it is
a reasonable approximation (as assumed here) to associate fragmentation
with a given, history independent, $\alpha$ value. More serious is 
the argument of Johnson and Gammie (2003) that a given $\alpha$ (or
cooling timescale) threshold is unreliable in the case that the cooling
timescale is a strong function of temperature: they showed  in
the regime where dust sublimes (and the opacity is a very steeply decreasing
function of temperature) that the onset of fragmentation did not
match their expectations based on the cooling timescale implied by  their
initial conditions. The problem in this case hinges on the initial guess
that one makes about the value of the Toomre Q parameter in the self-regulated
state: if the cooling rate is highly temperature dependent, then
a small adjustment in $Q$ from that imposed initially can result in a large
change in the cooling rate and thus a different fragmentation outcome
from that expected.  In practice this means that there are large
errorbars surrounding the location of the fragmentation boundary in the dust
sublimation regime. We will however see (Figure 1) that this corresponds to
rather high accretion rates that are not normally encountered in low
mass stars. 

\section{Thermal equilibrium solutions}

  Through our calculation of $\nu(\Sigma,R)$, we can assign to each
pair of parameters ($\Sigma$ and $R$)  the accretion rate that would
correspond to this solution if the disc were in a steady state. In the
case of a disc where the torque vanishes at the origin, 
we have 

\begin{equation}
\dot M_{ss} = 3 \pi \nu \Sigma
\end{equation}

 We emphasise that our local solutions do not require the disc to be
in a steady state and  that we use $\dot M_{ss}$ only as a convenient way of
parameterising our solutions: in general,  the actual accretion rate is
related to $\dot M_{ss}$ via: 

\begin{equation}
\dot M = \dot M_{ss} + 2R {{\partial \dot M_{ss}}\over{\partial R}}
\end{equation}

\begin{figure}
\vspace{2pt}
\epsfig{file=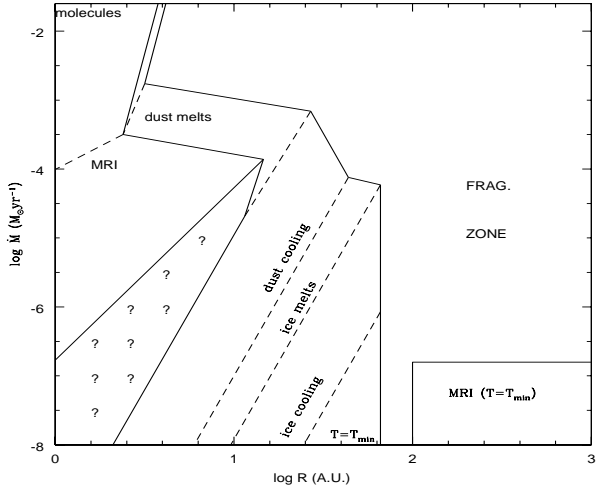,width=8.5cm,height=7.cm}
\caption{Regimes in the plane of steady state accretion rate versus
radius in the case of a disc surrounding an object of mass $1 M_\odot$.
The non-fragmenting, self-regulated self-gravitating regime lies between
the bold lines. See text for details.} 
\end{figure}

We can now (Figure 1) classify all points in the
parameter space of radius versus steady state accretion rate
and draw regime boundaries in this space (see the Appendix for analytic
expressions for solutions in the various regimes and for the location
of regime boundaries).

  The results of Figure 1 can be summed up as follows. 
The regions between the bold lines represent the range of
parameter space for which the disc is expected to be in the
self-regulated, self-gravitating regime: in other words,
angular momentum transport is dominated by gravitational
torques, but the disc is not expected to fragment. For most
of this region, the thermal input to the gas is also dominated
by heating associated with gravitational modes. The exception to this
is the lower right of this region (low accretion rates, large
radius) where the gas is mainly heated by an assumed interstellar
radiation field. Although the gas is assumed to be isothermal at
$10$ K in this region, it is deemed not to fragment because the cooling
timescale is long enough that perturbations behave quasi-adiabatically.

   The topology of the boundaries of this region is set both by the
fragmentation boundary (right hand boundary) and the interface with regions
where the  MRI dominates the angular momentum transfer (left hand boundary).  A notable feature of the former
is the vertical boundary at $\sim 70$ A.U. which coincides with regions
of the disc where the opacity is dominated by ice grains. In this
case $\kappa \propto T^2$, and therefore, for an optically thick disc
in a state of marginal gravitational instability, the cooling timescale
is simply a function of radius (i.e. independent of accretion rate, provided
one remains in the ice cooling regime). Consequently, the fragmentation
boundary (which can be cast either in terms of a critical value
of $\alpha$ or, equivalently, in terms of the ratio of cooling timescale to
dynamical timescale) is encountered at a fixed radius, a feature
first noted by Rafikov 2005  (see also Matzner \& Levin 2005, 
Stamatellos et al 2007). 
Outside this
radius, therefore, a self-gravitating disc is {\it always} subject to
fragmentation, 
regardless of the accretion rate. {\footnote{This also applies, 
for self-gravitating discs at $> 70 $ A.U.,  in the case that  the disc
temperature is set by external
irradiation. , since the fact that the cooling timescale is short compared
implies that perturbations behave approximately isothermally. We note that
the relationship between cooling timescale and $\alpha$ (equation 2)
no longer applies
when the disc heating is not dominated by the gravitational modes, and that
in this regime the disc will therefore fragment even at accretion rates
corresponding to very low $\alpha$ values.}} 

  We also see that beyond the dead zone (i.e. at $r > R_{dead}$ where
we have assumed $R_{dead} = 100$ A.U. in Figure 1) there is  a
region where the disc is non-self gravitating, with angular momentum
transport effected by the MRI. 
We can readily compute the maximum value of the steady state
accretion rate that is possible in this MRI region by noting
that equations (1), (6) and (7) can be re-cast as:

\begin{equation}
\dot M_{ss} = {{3 \alpha c_s^3}\over{GQ}}
\end{equation}

Thus for a disc that is isothermal at $T=T_{min}=10$K and
$\alpha=\alpha_{MRI}= 0.01$ we see that the maximum value of
$\dot M_{ss}$ for which the disc is self-consistently non-self gravitating
is a few times $10^{-7} M_\odot$ yr$^{-1}$. There is thus an island
of parameter space in the lower right of the diagram where fragmentation is
not expected.

Turning again to the self-regulated self-gravitating region of the
disc, we see that
at the lowest accretion rates the disc is isothermal with
temperature $T=T_{min}=10$K, but as the accretion rate is
increased, the thermal equilibrium temperature rises
above $T_{min}$, with dissipative heating balancing
optically thick radiative cooling, and  opacity provided
successively by  ice and dust. Only at very high accretion
rates ($> 10^{-4} M_\odot$ yr$^{-1}$) is the implied $\alpha$
value in excess of $\alpha_{frag}$. We therefore confirm
that fragmentation is not expected in the inner regions of discs
around low mass stars unless they are subject to extremely high
infall rates, a result that is consistent with radiation hydrodynamical
simulations of self-gravitating discs (Boley et al 2006,
Stamatellos \& Whitworth 2008) 
Inward of $33$ A.U., the disc becomes hot enough for dust to
start subliming before the point at which it fragments, and so
there is a self-gravitating regime where the opacity is a
steeply decreasing function of temperature and where the disc
is consequently nearly isothermal. Inward of $11$ A.U., 
there is a region where the mid-plane effective temperature is
$> 1000$K and where the value of $\alpha$ delivered by a
self-gravitating solution is less than $\alpha_{MRI}$:
consequently we expect this region to be MRI dominated at 
intermediate accretion rates, but self-gravitating at higher
or lower accretion rates. (Note that as discussed in 2.3 above, there is a 
region of parameter space where we cannot find self-consistent
solutions, because the self-gravitating solution has $T> 1000$K and
the MRI solution has $T<1000$K). 

  We stress that some of the above conclusions are dependent on our
assumptions about the efficacy of the MRI (i.e. our assumed values
of $R_{dead}$ and $\alpha_{MRI}$). For example, if
we assumed $R_{dead} < 70$ A.U., then the isothermal, MRI dominated region
in the lower right would join onto the  self-regulated self-gravitating 
disc at smaller radius and it would
then be possible for a disc fed at a low rate to avoid fragmentation altogether. 
Likewise, if $\alpha_{MRI}$ was lower than we have assumed, then the thermally
ionised MRI regime at a few A.U. would become self-gravitating at lower
accretion rate than shown in the Figures. 

 \begin{figure}
 \vspace{2pt}
 \epsfig{file=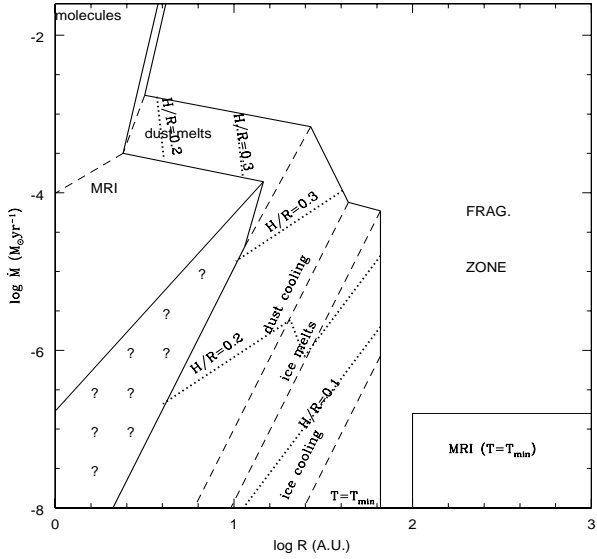,width=8.5cm,height=8.cm}
 \caption{Contours of disc equal axis ratio ($H/R$) in the case
 of a central object of mass $1 M_\odot$}
 \end{figure}
 
 \begin{figure}
 \vspace{2pt}
 \epsfig{file=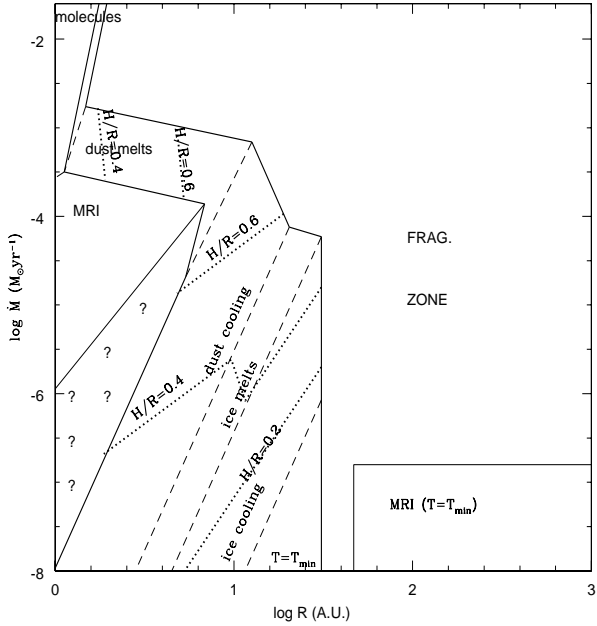,width=8.5cm,height=9.cm}
 \caption{Contours of disc equal axis ratio ($H/R$) in the case
 of a central object of mass $0.1 M_\odot$}
 \end{figure}

  Figure 2 depicts contours of $H/R$ in the $\dot M_{ss},r$ plane for
a solar mass central object and 
demonstrates that we are in the regime of lowish $H/R$ over most of the
self-regulated regime, thus justifying, a posteriori, the pseudo-viscous
approach (see discussion in Section 1). We note that since all disc quantities
(such as scale height H) relate to radius purely as a function of  $\Omega$, 
the  Keplerian
angular frequency, a given value of $H$ (at fixed $\dot M_{ss}$) is associated
with a value of R that scales with central object mass as $M^{1/3}$; therefore,
the corresponding diagram for a lower mass star is shifted to smaller radii
(by a factor $M^{1/3}$) and each $H/R$ contour is increased by a factor
$M^{-1/3}$. Figure 3 demonstrates that, for given accretion rates, 
self-regulated discs around low mass stars are geometrically thicker and
thus less well described by a pseudo-viscous (local description). Conversely
for higher mass stars (and supermassive central objects) discs in the
self-regulated regime (i.e. those that do not fragment) are geometrically
thinner than in the solar mass case (Figure 2a).

\begin{figure}
\vspace{2pt}
\epsfig{file=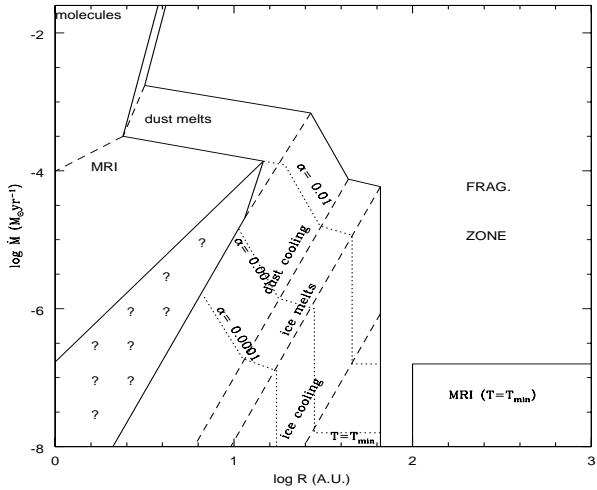,width=8.5cm,height=7.cm}
\caption{Contours of equal viscous $\alpha$ parameter}
\end{figure}

  Figure 4 depicts contours of $\alpha$ in the self-regulated regime.
These contours are, by construction, parallel to the fragmentation boundary
in all regimes where gravitational heating predominates. We draw attention to
the fact that  over a wide  range of parameter space, the 
$\alpha$ values delivered by self-gravitating discs are very low; this
presents a challenge to hydrodynamical modeling of disc formation since
for $\alpha$ less than about   $0.01$,  the angular momentum transport in  current codes is 
dominated by numerical viscosity.

 \begin{figure}
 \vspace{2pt}
 \epsfig{file=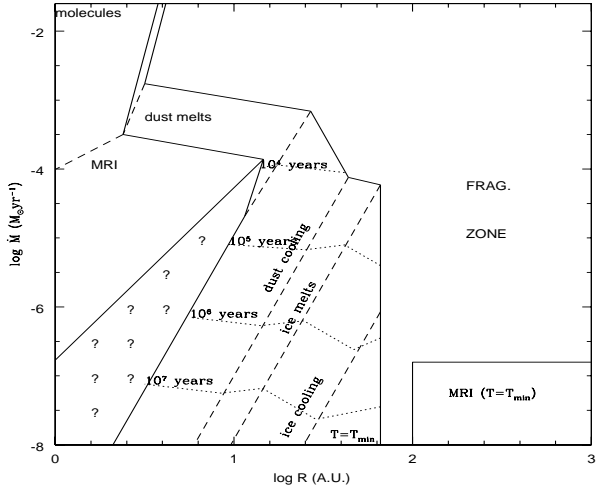,width=8.5cm,height=7.cm}
 \caption{Contours of equal viscous timescale in the case of
 a central object of mass $1 M_\odot$}
 \end{figure}

 \begin{figure}
 \vspace{2pt}
 \epsfig{file=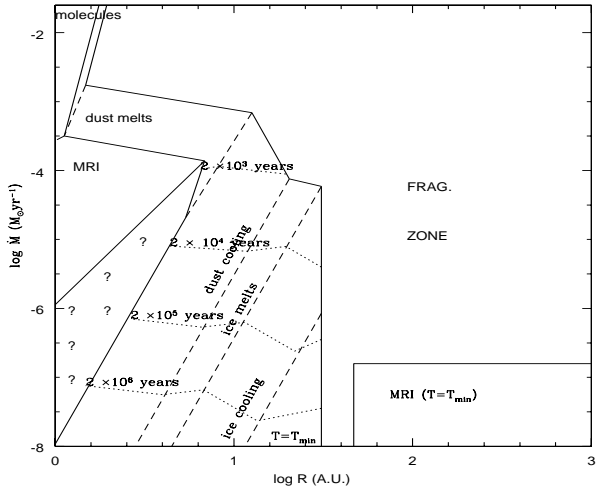,width=8.5cm,height=7.cm}
 \caption{Contours of equal viscous timescale in the case of
 a central object of mass $0.1 M_\odot$}
 \end{figure}

   Figure 5 and 6 depict contours of constant viscous timescale ($t_\nu = R^2/\nu$) for objects of central mass $1 M_\odot$ and $0.1 M_\odot$ respectively.
 This represents the characteristic timescale for mass and angular momentum
 transport in the disc and will be used in our subsequent discussion of
 disc secular evolution. We draw attention to the fact that these timescales
are many orders of magnitude in excess of the dynamical timescales at
these  radii and that it is therefore appropriate to discuss the secular
evolution of such discs. 

\section{Applications}
\subsection {Comparison with hydrodynamical infall models}
Our a posteriori justification of the pseudo-viscous approach
(see Figure 2), means that we can use the  solutions presented in
the Appendix to compute the secular evolution of discs in the process
of assembly from a collapsing core.
This involves integration of the the time-dependent viscous
diffusion equation:

\begin{equation}
 {{\partial \Sigma}\over {\partial t}} = {3\over
R}{{\partial} \over{ \partial R}}\biggl(R^{1\over2}{{\partial}
\over{\partial R}} ( R^{1\over2}\nu \Sigma)\biggr) + \dot \Sigma_{inf}
\end{equation}

where $\dot \Sigma_{inf}$ is the parameterised infall rate
(mass per unit area per unit
time).  At given $\Sigma$ and $R$, the value of
$\nu \Sigma$ required for the solution of equation (10) can be
simply derived from the $\Sigma (\dot M_{ss}, R)$ solutions given
in the Appendix, noting the relationship between $\dot M_{ss}$
and $\nu \Sigma$ contained in equation (7).
We have undertaken some trial experiments of this sort which  confirm 
the statements we make below on analytic grounds. Nevertheless, the
real utility of this approach will be its use in analysing the results
of hydrodynamical simulations of collapsing cores, where one will
obtain $\dot \Sigma_{inf}$ numerically and can thus compare the secular
evolution of the simulation with that predicted by a pseudo-viscous
treatment. We emphasise the need for simulators to take note of 
Figure 4, in order that they avoid regimes where the expected
$\alpha$ delivered by gravitational instability is less than that
associated with the numerical viscosity in their codes.

\subsection{Fragmentation}

 In self-gravitating discs that are optically thick, with opacity
provided by ice grains, the cooling timescale is a function of radius
only (Rafikov 2005)  and is independent of
temperature. This implies a particular radial location $r_{frag}$
for disc fragmentation in this regime, independent of accretion rate.
As has been  
noted by several previous authors (Rafikov 2005, Matzner \& Levin 2005,
Stamatellos et al 2007) $r_{frag}$ is around $70$ A.U. for solar mass stars; inward
of $r_{frag}$, fragmentation can only be expected for  high accretion 
rates ($> 10^{-4} M_\odot$ yr$^{-1}$). 

 Obviously, therefore, one expects   disc fragmentation 
in the case of cores whose specific angular momentum is high
enough for significant infall beyond $\sim 70$ A.U.. In terms of
the factor $\beta_J$ (the ratio of the rotational kinetic energy of
the core to its break-up value) this implies $\beta_J > 0.01$ for solar
mass cores collapsing from a Jeans scale at temperature $\sim 10$K. However,
our results furthermore imply that fragmentation at $r_{frag}$ is 
inevitable for {\it any} plausible initial core rotation rate, because of
the possibility of outward transfer of angular momentum (and mass) even
in the case where the maximum radius of infall ($r_{inf}$) 
is restricted to radii $<<r_{frag})$.
All that is required in this case is that there is enough angular
momentum and mass in the infalling material for it to be self-gravitating
at $> r_{frag}$, following radial redistribution by gravitational
torques. Since the specific angular
momentum in a Keplerian disc scales as $R^{1/2}$), then we require that
if a disc spreads so that mass $M_{frag}$ ends up at radius larger
than $r_{frag}$, then we must have $M_{disc} r_{inf}^{1/2} >
M_{frag} r_{frag}^{1/2}$. Given the typical parameters of self-gravitating
discs at $r> r_{frag}$, we require that $M_{frag}$ is at least around
$10 \%$ of the central object mass. Since the entire central object mass
is initially in the disc, we then require that the disc initial
infall radius is at
least around $0.01 r_{frag}$. This radius is extremely small (less than
an A.U.) and would correspond to an implausibly low initial core rotation
rate ($\beta_J < 10^{-4}$).
We thus conclude that, because of the
efficient outward angular momentum transfer in self-gravitating discs,
disc fragmentation at $> r_{frag}$ is inevitable in just about any core 
with a realistic  initial
angular momentum content. This conclusion is not
substantially changed if we relax the assumption that the MRI is effective
only beyond $100$ A.U.. If instead we postulate that the disc is MRI active
down to radius $r_{frag}$ then we could in principle avoid fragmentation if
the MRI took over as an angular momentum transfer mechanism once material
reached $r_{frag}$. We however see from Figure 1 that this places an upper
limit on the rate  at which mass flows outwards at $r_{frag}$.   
Our numerical experiments, involving the integration of equation (10) for a
variety of disc infall histories, indicate that this condition is never
met unless the  rate of infall is very low ($< 10^{-7} M_\odot$ yr$^{-1}$:
see discussion following equation (9)).
Collapse models for low mass cores (e.g. Vorobyuov \& Basu 2005 and
references therein) however
suggest rates
that are an order of magnitude higher than this and these higher values
($\sim 10^{-6} M_\odot $ yr$^{-1}$) are corroborated by modeling of line
profiles in such cores (e.g. Tafalla et al 2000).  

  The two important elements to emerge from this discussion are
i) the thermodynamics of self gravitating discs only permits
fragmentation at $r > r_{frag} \sim 70 M^{1/3}$ A.U. and ii) virtually
any core with non-negligible angular momentum will produce a disc that
will expand, due to the action of gravitational torques, to radii
$> r_{frag}$. {\footnote{This latter conclusion assumes that there
is not substantial loss of angular momentum from the disc as
a result of winds or outflows at this stage}}
Both these points are  relevant to binary
star formation and it is tempting to make the
association between $r_{frag}$ and 
the median binary 
separation (which is observed to be similar to this 
e.g. Duquennoy \& Mayor 1991, Fischer \& Marcy 1992).

  There are several aspects of this model which have not been captured by
simulations of binary star formation to date. First of all, the
model  suggests
that fragmentation should be the norm, since virtually all cores are
likely to have enough angular momentum for their resulting
discs to expand to $r_{frag}$. In this scenario, therefore, the initial
location of fragmentation does not depend, in the limit of low core
angular  momentum on the actual value of this angular momentum, 
since the characteristic fragmentation radius is instead
set by the thermodynamics of ice dominated cooling. Secondly, where
the disc expands out to $r_{frag}$ before fragmentation occurs, the
material involved in the fragmentation now has  higher specific angular
momentum than any other material in the core. This is important for
the mass ratio, $q$, of the resulting binary: not only  is the companion
initially of rather low mass (being composed of the high angular momentum
tail of material that has attained radii $> r_{frag}$) but it will not be
driven to higher $q$ by subsequent accretion of higher angular momentum
material. 

  This therefore avoids a well known problem of binary star
formation, i.e. the tendency of binaries to end up with nearly equal
mass ratios,  in contrast to the observed situation in which low mass ratios
are abundant (Duquennoy \& Mayor 1991, Halbwachs et al 1993, Prato et al
2002). This tendency for simulations to produce nearly equal mass pairs
is driven by  the late accretion of high specific angular
momentum material onto the proto-binary pair, which favours preferential
accretion onto the secondary (Whitworth et al 1995, Bate \& Bonnell 1997, Bate 2000,  Delgado et al
2004). In most previous calculations,
which employ an isothermal equation of state for all but the densest
regions of the disc, fragmentation ensues as soon as the disc is formed
and therefore the fragment cannot avoid  interacting with high specific angular
momentum material that continues to fall in from the outer regions of
the core. The new element here is that fragmentation is delayed until
the disc has re-expanded out to $r_{frag}$. Figure 4 shows that this occurs
on a timescale of $\sim 10^5-10^6$ years, which is somewhat longer than the
free fall timescale of the collapsing core. {\footnote {It should be noted that the recent simulations
of Attwood et al 2009 also demonstrate delayed fragmentation: in this case
the delay is instead due to the time required to assemble a Toomre unstable
disc, which is a shorter timescale ($1-3 \times 10^4$ years) than the time
required for angular momentum transport discussed here. This shorter time
is now no longer much longer than the core infall timescale, which probably
explains why these simulations still show a tendency towards growth
of $q$ towards unity through accretion.}}Although this scenario
has to be substantiated by hydrodynamical calculations that incorporate
the necessary physics, this {\it delayed fragmentation} provides
a promising mechanism for the formation of low q binaries.

\section{Conclusions}

  Our derivation of analytic expressions for the properties of
self-gravitating self-regulated
discs has allowed us to derive useful diagrams illustrating different
physical regimes as a function of steady state accretion rate 
and radius (Figures
1-4). These delineate the fragmentation boundary and illustrate that
for all but the lowest mass stars, one expects  non-fragmenting discs
to be well
described by the local (pseudo-viscous) approach employed here. These plots
also demonstrate,  as noted by several previous authors, that fragmentation
is only expected at radius $> r_{frag} \sim 70 (M/M_\odot)^{1/3}$ A.U. provided
that infall from the parent core is at less than $\sim 10^{-4} M_\odot$
yr $^{-1}$. Thus it is only in high mass cores ($> 10 M_\odot$, where such
high infall rates are deduced (Cesaroni et al 2007)) that one would expect fragmentation at smaller
radius. 
On the other hand, fragmentation at $> r_{frag}$ is pretty much inevitable
for any plausible initial core rotation rates: the only scenario in which
such fragmentation could be avoided is in the case both that the MRI
extends in to $r_{frag}$ and if the mass infall rate from the parent
core is very low ($<10^{-7} M_\odot$ yr$^{-1}$).

  We point out that our analytic expressions provide a useful framework
for comparison with future hydrodynamic collapse calculations that
incorporate the necessary cooling physics. In particular, Figure 4
raises a cautionary note about the very low viscous $\alpha$ values to
be expected in some regions of parameter space for self-gravitating,
self-regulated discs and shows that care must be taken that calculations
are not instead dominated by the effect of numerical viscosity. We also
point out that our steady state expressions (listed in the Appendix)
can be readily re-arranged so as to find the effective viscosity of
self-gravitating discs as a function of surface density and radius and
thus enable the secular evolution of the disc to be computed using
the viscous diffusion equation. Such calculations may then be usefully
compared with the results of hydrodynamical modeling. 

  We mainly apply our results to the issue of binary star formation,
pointing out that, fortuitously or not, $r_{frag}$ is close to the
median binary separation. We show that the unlikelihood of fragmentation
within $r_{frag}$ offers the possibility, in low angular momentum
cores, of {\it delayed} fragmentation, whereby the disc is assembled
at small radii and fragmentation only then ensues later, one the disc has
spread outwards, through the action of gravitational torques, to $>
r_{frag}$. This delayed fragmentation, that would not be seen in models
that employ a mainly isothermal equation of state in the disc,  has a
distinct advantage when it comes to reproducing observed binary statistics.
In this case, fragmentation is likely to occur {\it after} the bulk of
the parent core has fallen in, and thus the fragment would avoid the
accretion of infalling material which, in current models, drives binary mass
ratios to high values. Since observations are unambiguous in requiring a
large population of binaries (with separations of $10-100$s of A.U.) that
have low mass ratios ($q < 0.5$), this is an outstanding shortcoming of
current models. We propose however that the different thermal regimes
at $r < $ and $> r_{frag}$ offer a way to solve the problem of the
creation of low $q$ binaries.  

\section{Acknowledgments}
I am grateful to Giuseppe Lodato for providing useful comments on
an earlier draft and to the referee, Anthony Whitworth, for constructive
criticism. 

\section {Appendix}

In order to derive `standard self-regulated' solutions in thermal
equilibrium (see Section 2),   we solve equations (1)-(7), 
together with parameterisations for the 
opacity taken from Bell
and Lin 1994. For Rosseland mean opacity parameterised in the
form $\kappa = \kappa_0 \rho^a T^b$, the thermal equilibrium 
solution for optically thick cooling is given by

\begin{equation}
\nu \Sigma = C \Sigma^{7-2b} R^{15-3b+3a}
\end{equation}

where
\begin{equation}
C = {{G^{3-b} \sigma}\over{M_*^{5-b+a} \kappa_0 {\cal R}^{4-b}}}
\end{equation}

where $\cal R$ is the gas constant, such that $c_s^2 = \cal R T$.

In the opacity regimes  relevant to our calculations, we
have $\kappa_0 = 2 \times 10^{-4}, a=0, b=2$ (for opacity provided
by ice grains), $\kappa_0 = 2 \times 10^{16}, a=0, b=-7$ (for
the region where ice grains sublime) and $\kappa_0 = 0.1, a=0,
b=0.5$ for the region where dust grains dominate the opacity.

 These opacity regimes are therefore associated with the following
optically thick, self-regulated solutions (for $M_*=1 M_\odot$):
\bigskip

A. Ice grains.

\begin{equation}
\Sigma = 14 \dot M_{-6}^{1/3} R_{100}^{-3} {\rm{g cm}}^{-2}
\end{equation}

\begin{equation}
T = 3.2 \dot M_{-6}^{2/3}R_{100}^{-3} \rm K
\end{equation}

\begin{equation}
\alpha = 0.4 R_{100}^{9/2}
\end{equation}

\begin{equation}
H/R = 0.055 \dot M_{-6}^{1/3} R_{100}^{-1}
\end{equation}

\begin{equation}
t_{\nu} = 1.7 \times 10^5 \dot M_{-6}^{-2/3}  R_{100}^{-1} {\rm years}
\end{equation}

where $R_{100}$ is radius normalised to $100$ A.U. and
$\dot M_{-6}$ is the steady state accretion rate (equation 8)
normalised to $10^{-6} M_\odot$ yr$^{-1}$.  

\bigskip
B. Ice grains sublime 

\begin{equation}
\Sigma = 75 \dot M_{-6}^{1/21} R_{100}^{-12/7} {\rm{g cm}}^{-2}
\end{equation}

\begin{equation}
T = 95. \dot M_{-6}^{2/21} R_{100}^{-3/7} \rm K
\end{equation}

\begin{equation}
\alpha= 2.5 \times 10^{-3} R_{100}^{9/14} \dot M_{-6}^{6/7}
\end{equation}

\begin{equation}
H/R = 0.30 \dot M_{-6}^{1/21} R_{100}^{2/7}
\end{equation}

\begin{equation}
t_\nu = 9.4 \times 10^{5} \dot M_{-6}^{-20/21} R_{100}^{2/7} {\rm years}
\end{equation}

\bigskip
C. Dust grains

\begin{equation}
\Sigma = 29 \dot M_{-6}^{1/6} R_{100}^{-9/4} {\rm{g cm}}^{-2}
\end{equation}

\begin{equation}
T = 14 \dot M_{-6}^{1/3} R_{100}^{-1.5} \rm K
\end{equation}

\begin{equation}
\alpha= 4.4 \times 10^{-2} R_{100}^{9/4} \dot M_{-6}^{1/2}
\end{equation}

\begin{equation}
H/R = 0.11 \dot M_{-6}^{1/6} R_{100}^{-1/4}
\end{equation}

\begin{equation}
t_\nu = 3.4 \times 10^5 \dot M_{-6}^{-5/6} R_{100}^{-1/4} {\rm years}
\end{equation}

D. Dust grains sublime
\begin{equation}
\Sigma = 170  \dot M_{-6}^{0.018} R_{100}^{-1.64} {\rm{g cm}}^{-2}
\end{equation}

\begin{equation}
T = 500  \dot M_{0.036}^{1/3} R_{100}^{-0.27} \rm K
\end{equation}

\begin{equation}
\alpha= 2.1 \times 10^{-4} R_{100}^{0.41} \dot M_{-6}^{0.945}
\end{equation}

\begin{equation}
H/R = 0.68 \dot M_{-6}^{0.018} R_{100}^{0.36}
\end{equation}

\begin{equation}
t_\nu = 2.1 \times 10^6 \dot M_{-6}^{-0.82} R_{100}^{0.36} {\rm years}
\end{equation}

\bigskip
These solutions are continuous at the zone boundaries, which are
located at

\begin{equation}
R_{ice melt} = 27 \dot M_{-6}^{2/9} \rm{A.U.}
\end{equation}

(where ice grains sublime)  

\begin{equation}
R_{dust} = 17 \dot M_{-6}^{2/9} \rm{A.U.}
\end{equation}
 
(where dust grains dominate the opacity) and 

\begin{equation}
R_{dust melt} = 5.5 \dot M_{-6}^{0.24} \rm{A.U.}
\end{equation}

where dust grains sublime.
In addition, the disc is isothermal (with $T=T_{min}=10$K) for
$R > R_{iso}$ where

\begin{equation}
R_{iso} = 70 \dot M_{-6}^{2/9} {\rm A.U.}
\end{equation}

Throughout the isothermal regime, the disc properties are described by:

\begin{equation}
\Sigma = 25 R_{100}^{-1.5} {\rm g cm}^{-2}
\end{equation}
\begin{equation}
\alpha = 0.06 \dot M_{-6}
\end{equation}

\begin{equation}
H/R = 0.1 R_{100}^{0.5}
\end{equation}

and

\begin{equation}
t_\nu = 3 \times 10^5 R_{100}^{0.5} \dot M_{-6}^{-1} {\rm years}
\end{equation}


\end{document}